\documentclass[twocolumn,english]{article}
\usepackage[T1]{fontenc}
\usepackage[latin9]{inputenc}
\usepackage{textcomp}
\usepackage{amsbsy}
\usepackage{amstext}
\usepackage{amssymb}
\usepackage{graphicx}
\usepackage{esint}

\makeatletter
\newcommand{\lyxaddress}[1]{
\par {\raggedright #1
\vspace{1.4em}
\noindent\par}
}

\makeatother

\usepackage{babel}
\begin{document}
\twocolumn[\begin{@twocolumnfalse}

\title{A variational formulation for dissipative fluids with interfaces\linebreak{}
in an inhomogeneous temperature field }

\author{Hiroki Fukagawa\textsuperscript{1,2,3,{*}}, Chun Liu\textsuperscript{2},
and Takeshi Tsuji\textsuperscript{1}}

\maketitle

\lyxaddress{\begin{center}
\textsuperscript{1}International Institute for Carbon-Neutral Energy
Research (I2CNER), Kyushu University, Fukuoka 819-0395, Japan\linebreak{}
\textsuperscript{2}Department of Mathematics, Penn State University,
University Park, PA 16802, USA\linebreak{}
\textsuperscript{3}School of Fundamental Science and Technology,
Keio University, Yokohama 223-8522, Japan
\par\end{center}}
\begin{abstract}
We derive the equations governing the motion of fluids with interfaces
in an inhomogeneous temperature filed, by employing a variational
principle. Generally, the Lagrangian of a fluid is given by the kinetic
energy density minus the internal energy density. The dynamics also
obeys the equation of entropy. Then the necessary condition for minimizing
an action with subject to the constraint of entropy yields the equation
of motion. In this way, this method provides the equations of a fluid
when the kinetic and internal energies and the equation of entropy
are given. However, it is sometimes to know the proper equation of
entropy. Our main purpose of this article is to determine it by using
the three requirements, which are a generalization of Noether's Theorem,
the second law of thermodynamics, and well-posedness. To illustrate
this approach, we investigate several phenomena in an inhomogeneous
temperature field. In the case of vaporization, diffusion and the
rotation of a chiral liquid crystals, we clarify the cross effects
between the entropy flux and these phenomena via the internal energy.
\end{abstract}
\end{@twocolumnfalse}]
\renewcommand\thefootnote{*}
\footnotetext[1]{fukagawa.hiroki.609@m.kyushu-u.ac.jp}

\section{Introduction\label{sec:Introduction}}

The dynamics of a fluid can be divided into a kinetic part and a thermodynamics
part. The kinetic part is characterized by the equations of the conservation
laws for mass, energy, momentum, and angular momentum. On the other
hand, the thermodynamics part obeys the second law of thermodynamics,
which states that the total entropy of a closed system is always increasing.
The equations of motion and the equation of entropy should be consistent
with these conservation laws and the second law of thermodynamics.
However, it is sometimes difficult to know the both proper equations
for fluids with heat transfer, such as two-phase flows, two-component
fluids, and liquid crystals. Thus we need a procedure for deriving
these equations systematically.

On the assumption of constant temperature and low kinetic energy,
the total Helmholtz free energy of a closed isolated system tends
to decrease because of the conservation law of total energy and the
second law of thermodynamics\cite{adkins1983equilibrium}\emph{.}
Various models based on the Helmholtz free energy cannot be applied
to the fluids in an inhomogeneous temperature field by themselves,
in spite of widely used in the fields of complex fluids\cite{JDvanderWaals,cahn1958free,doi:10.1143/JPSJ.78.052001,Lisin199755,Lisin1999327,PhysRevE.86.031703},
Then we have to use heuristic methods to combine thermodynamics with
these models in order to apply to these phenomena\cite{Leslie19111968,de1993physics}.

In this article, we propose a new formulation without using the Helmholtz
free energy and the heuristic methods. Originally, the physical system
without thermodynamics can be expressed by the variational principle
in analytical dynamics. This principle states that the realized motion
minimizes an functional, which is called an action in physics. In
our previous works and related work\cite{Fukagawa01052012,Fukagawa01092010,fukagawa2012,hyon2010energetic,liu2009introduction,2014arXiv1407.1035J,aoki2014constraint},
the equations of motion for complex fluids without interfaces in an
inhomogeneous temperature field are derived from this variational
principle with the aid of the constraint of entropy. However, it is
sometimes difficult to know it, especially in the case of the fluids
with interfaces. The main purpose of this article is to obtain the
proper constraint of entropy. This article is organized as follows.
We review the optimized control theory, which is considered as the
generalization of the variational principle used in physics in Sec.~\ref{sub:Lagrangian}.
We explain the basic concept of our new method in Sec.~\ref{sec:theory}.
We apply it to the realistic fluid, such as a Newtonian fluid, and
various fluids with interfaces in Sec.~\ref{sec:Advanced-theory-for}.
We summarize our formulation in Sec.~\ref{sec:Discussion}.

\section{Optimal control theory\label{sub:Lagrangian}}

Here, we give a short review of the optimal control theory of a nonholonomic
system\cite{bloch2003nonholonomic}. The degree of freedom is defined
as the number of the state variables needed to express the dynamics
of a system at least. Let $\boldsymbol{q}=(q_{1,}\cdots q_{n})$ be
state variables except entropy $s$. The dynamics is described as
a trajectory in the configuration space $(\boldsymbol{q},s,t)$, and
restricted by the constraint, 
\begin{equation}
Tds+f_{i}dq_{i}+Qdt=0,\label{eq:nonholo}
\end{equation}
where $T$, $\boldsymbol{f}=(f_{1,}\cdots f_{n})$, and $Q$ are coefficients.
Here we use $d$ for the exterior derivative. The constraint (\ref{eq:nonholo})
is a nonholonomic constraint, i.e.,  cannot be expressed as a function:
$U(\boldsymbol{q},s,t)=0$, because the entropy $s$ cannot determined
only by other variables $\boldsymbol{q}$ and time $t$. In terms
of optimized control theory, the physical system with dissipation
is considered as a nonholonomic system, and the equation of the motion
is regard as a resultant equation of the optimized control\cite{pontryagin1987mathematical,free,bloch2003nonholonomic},
which minimizes a value functional (an action) with subject to Eq.~(\ref{eq:nonholo}).
The time evolution of $\boldsymbol{q}$ is given by 
\begin{equation}
\frac{d\boldsymbol{q}}{dt}=F(\boldsymbol{q},\boldsymbol{u}).\label{eq:cont}
\end{equation}
Here the vector $\boldsymbol{u}$ is called control parameter because
the function $F(\boldsymbol{q},\boldsymbol{u})$ determines the dynamics.
In control theory,  $\int_{t_{{\rm init}}}^{t_{{\rm fin}}}dt\ L(\boldsymbol{q},\boldsymbol{u},s)$
is called evaluation functional. By the method of undetermined multipliers,
the functional to be minimized is
\begin{equation}
I[\boldsymbol{q},\boldsymbol{p},\boldsymbol{u},s]\equiv\int_{t_{{\rm init}}}^{t_{{\rm fin}}}dt\ \tilde{L},\label{eq:Iqpu-3}
\end{equation}
where $\tilde{L}$ is defined as
\begin{eqnarray}
\tilde{L} & \!\!\!\!\equiv\!\!\!\! & L(\boldsymbol{q},\boldsymbol{u},s)+\boldsymbol{p}\cdot\left(\!\frac{d\boldsymbol{q}}{dt}\text{\textminus}F(\boldsymbol{q},\boldsymbol{u})\!\right)\nonumber \\
 & \!\!\!\!=\!\!\!\! & -\tilde{H}(\boldsymbol{q},\boldsymbol{p},\boldsymbol{u},s)+\boldsymbol{p}\cdot\frac{d\boldsymbol{q}}{dt},\label{eq:tildL-3}
\end{eqnarray}
with the aid of undetermined multipliers $\boldsymbol{p}=(p_{1},\cdots,p_{n})$.
We define $\tilde{H}$ as $\tilde{H}\equiv-L+\boldsymbol{p}\cdot\boldsymbol{F}$.
The necessary condition of minimizing the action (\ref{eq:Iqpu-3})
with subject to Eq.~(\ref{eq:nonholo}) is given by
\begin{eqnarray}
 &  & \!-\!\left(\!\frac{dp_{i}}{dt}\!+\!\frac{\partial\tilde{H}}{\partial q_{i}}\!-\! f_{i}\!\right)\! dq_{i}\!+\!\left(\!\frac{dq_{i}}{dt}\!-\!\frac{\partial\tilde{H}}{\partial p_{i}}\!\right)\! dp_{i}\nonumber \\
 &  & -\!\frac{\partial\tilde{H}}{\partial u_{i}}\! du_{i}\!-\!\left(\!\frac{\partial\tilde{H}}{\partial s}\!-\! T\!\right)\! ds\!\!=\!\boldsymbol{0}.\label{eq:Iqpu-4}
\end{eqnarray}
We show the details in App.~\ref{sec:differential-form}. Without
loss of generality, we can define

\begin{equation}
T\equiv-\frac{\partial L}{\partial s}=\frac{\partial\tilde{H}}{\partial s},\label{eq:s-1}
\end{equation}
and obtain
\begin{eqnarray}
0 & \approx & \frac{\partial\tilde{H}}{\partial u_{i}},\label{eq:u-1}\\
\frac{dq_{i}}{dt} & \approx & \frac{\partial\tilde{H}}{\partial p_{i}},\label{eq:q-1}\\
\frac{dp_{i}}{dt} & \approx & -\frac{\partial\tilde{H}}{\partial q_{i}}+f_{i}.\label{eq:p-1}
\end{eqnarray}
Here $\approx$ denotes weak equality, i.e., the left-hand side (lhs)
equals the right-hand side (rhs) when the constraint (\ref{eq:nonholo})
is satisfied. Let $H$ be the Hamiltonian defined by 
\begin{equation}
H(\boldsymbol{q},\boldsymbol{p},s)\equiv\tilde{H}(\boldsymbol{q},\boldsymbol{p},\boldsymbol{u}(\boldsymbol{q},\boldsymbol{p},s),s)\label{eq:Hamilton}
\end{equation}
where $\boldsymbol{u}(\boldsymbol{q},\boldsymbol{p},s)$ satisfies
Eq.~(\ref{eq:u-1}). Finally, we have
\begin{eqnarray}
\frac{ds}{dt} & = & -\frac{1}{T}\left(f_{i}\frac{dq_{i}}{dt}+Q\right),\label{eq:u-1-1}\\
\frac{dq_{i}}{dt} & = & \frac{\partial H}{\partial p_{i}},\label{eq:q-1-1}\\
\frac{dp_{i}}{dt} & = & -\frac{\partial H}{\partial q_{i}}+f_{i}.\label{eq:p-1-1}
\end{eqnarray}

\section{A proposed theory\label{sec:theory}}

Let us consider a physical system with dissipation, which can be considered
as a nonholonomical system. Physical systems often satisfy several
symmetries such as the Galilean symmetry, the translation symmetries
in time and space, and the rotational symmetry. Noether's theorem
connects the symmetries and the balance equations. We generalize it
to the nonholonomic system in Sec.~\ref{sub:Noether's-theorem-with}.
Next, we derive the constraint of entropy from the generalized Noether's
theorem with the aid of the second law of thermodynamics in Sec.~\ref{sub:Second-law-of},
and give a simple example in Sec.~\ref{sub:A-damped-oscillator}.

\subsection{Symmetries in a physical system\label{sub:Noether's-theorem-with}}

We use $\boldsymbol{r}\equiv(\boldsymbol{q},\boldsymbol{p},s)$ to
simplify the notation. Let us consider the infinitesimal transformation:
\begin{eqnarray}
\boldsymbol{r}' & = & \boldsymbol{r}+\alpha\boldsymbol{g_{r}},\label{eq:LR-1}\\
t' & = & t+\alpha g_{t},\label{eq:LT-1}
\end{eqnarray}
where $\alpha$ is an infinitesimal constant. The transformation~(\ref{eq:LR-1})
and (\ref{eq:LT-1}) defines symmetries of a system, if it moves an
optimized trajectory $\boldsymbol{r}^{*}(t)$ to other optimized trajectories
$\boldsymbol{r}'^{*}(t')$ of a nonholonomic system. We write $\boldsymbol{X}_{H}=\frac{dq_{i}}{dt}\partial_{q_{i}}+\frac{dp_{i}}{dt}\partial_{p_{i}}+\frac{ds}{dt}\partial_{s}$
for the vector field satisfying Eqs.~(\ref{eq:u-1-1})--(\ref{eq:p-1-1}).
The necessary and sufficient condition condition for a symmetry is
condition is that there is a vector field $\boldsymbol{X}_{G}=g_{q_{i}}\partial_{q_{i}}+g_{p_{i}}\partial_{p_{i}}+g_{s_{i}}\partial_{s}$
and a function $G$ such that
\begin{eqnarray}
\boldsymbol{0} & \!\!=\!\! & \left\{ \! g_{p_{i}}\!-\! g_{t}\!\left(\!-\!\frac{\partial H}{\partial q_{i}}\!+\! f\!\right)\!\right\} \! dq_{i}\!-\!\left(\! g_{q_{i}}\!-\! g_{t}\!\frac{\partial H}{\partial p_{i}}\!\right)\! dp_{i}\nonumber \\
 &  & -(\boldsymbol{X}_{G}H+g_{t}Q)dt+dG.\label{eq:dG-3}
\end{eqnarray}
See the derivation in App.~\ref{sec:differential-form-1}. Bellow,
we give some examples of $\boldsymbol{X}_{G}$ and $G$. Here and
below, we assume that the Hamiltonian $H(\boldsymbol{q},\boldsymbol{p},s)$
and the momentum map $G(\boldsymbol{q},\boldsymbol{p},s)$ does not
depend on time explicitly. Provided $g_{t}=0$, $g_{q_{i}}=\partial G(\boldsymbol{q},\boldsymbol{p})/\partial p_{i}$,
and $g_{p_{i}}=-\partial G(\boldsymbol{q},\boldsymbol{p})/\partial q_{i}$,
Eq.~(\ref{eq:dG-3}) becomes
\begin{equation}
\boldsymbol{0}=-(\boldsymbol{X}_{G}H)dt,\label{eq:LIE-1}
\end{equation}
and yields
\begin{eqnarray}
0 & = & \boldsymbol{X}_{G}H\nonumber \\
 & = & -\frac{\partial G}{\partial q_{i}}\frac{\partial H}{\partial p_{i}}+\frac{\partial G}{\partial p_{i}}\frac{\partial H}{\partial q_{i}}+g_{s}\frac{\partial H}{\partial s}\nonumber \\
 & = & -\frac{\partial G}{\partial q_{i}}\frac{\partial H}{\partial p_{i}}+\frac{\partial G}{\partial p_{i}}\left(\frac{\partial H}{\partial q_{i}}-f_{i}\right)\nonumber \\
 & = & -\frac{dG}{dt},\label{eq:balance-1}
\end{eqnarray}
if $\boldsymbol{X}_{G}$ satisfies 
\begin{equation}
Tg_{s}+f_{i}g_{q_{i}}+Qg_{t}=0.\label{eq:nonholo-1}
\end{equation}
A example of this symmetry is the space translation of a physical
system: $\boldsymbol{g}_{\boldsymbol{q}}=(1,\cdots,1)$, $g_{s}=g_{t}=0$,
and $G=\sum_{i}p_{i}$. The equations~(\ref{eq:balance-1}) and (\ref{eq:nonholo-1})
show that the total momentum $G=\sum_{i}p_{i}$ is conserved if we
have 
\begin{equation}
\sum_{i}f_{i}=0.\label{eq:ff}
\end{equation}
Next, let us consider the case of $\boldsymbol{X}_{G}=\boldsymbol{X}_{H}$,
$g_{t}=(1-\gamma)\partial_{t}$ and $G=\gamma H$, where $\gamma$
is a real constant number. This translation denotes a time translation.
The equation~(\ref{eq:dG-3}) yields 
\begin{equation}
\frac{dH}{dt}+Q=0.\label{eq:enecon}
\end{equation}
with the aid of Eqs.~(\ref{eq:q-1-1}) and (\ref{eq:p-1-1}). We
can interpret Eqs.~(\ref{eq:enecon}) as the energy conservation
law, where $H$ and $Q$ are the total energy and outflow of heat,
respectively. As seen in this section, Eq.~(\ref{eq:dG-3}) is a
generalization of Noether's theorem. If a nonholonomic system has
a symmetry, i.e, Eq.~(\ref{eq:dG-3}) is valid on a transformation,
we have a corresponding balance Eqs.~(\ref{eq:balance-1}) and (\ref{eq:enecon}).

\subsection{Second law of thermodynamics\label{sub:Second-law-of}}

Let us consider the environment. The constraint of entropy $S$ in
environment is given by 
\begin{equation}
T_{E}dS+Q_{E}=0,\label{eq:nonholo-2}
\end{equation}
and then the energy conservation law is
\begin{equation}
\frac{dH}{dt}+Q+Q_{E}=0,\label{eq:enecon-1}
\end{equation}
instead of Eq.~(\ref{eq:enecon}). By the definition of isolated
system, the total energy including environment $H$ is constant. It
yields $Q=-Q_{E}$ from Eq.~(\ref{eq:enecon-1}). Here, $Q$ and
$Q_{E}$ express heat flow. The second law of thermodynamics, which
states that the total entropy of isolated system increases, $\frac{d(s+S)}{dt}=\Theta>0$.
From Eqs.~(\ref{eq:nonholo}) and (\ref{eq:nonholo-2}), the dissipative
function denoting the entropy production per unit time is $\Theta\equiv-\frac{1}{T}\left(\boldsymbol{f}\cdot\frac{d\boldsymbol{q}}{dt}\right)+Q\left(\frac{1}{T_{E}}-\frac{1}{T}\right)$.
Remind that $d\boldsymbol{q}/dt$ is determined by $\boldsymbol{u}$
from Eq.~(\ref{eq:cont}), the dissipative function $\Theta$ is
a function of $\tilde{\boldsymbol{u}}=(\boldsymbol{u},\left(1/T_{E}-1/T\right))$.
Low degree approximation of $\Theta(\tilde{\boldsymbol{u}})$ is given
in the quadratic form:
\begin{equation}
\Theta(\tilde{\boldsymbol{u}})=\tilde{\boldsymbol{u}}^{t}M\tilde{\boldsymbol{u}}\geq0,\label{eq:quadro}
\end{equation}
where $M$ is symmetric coefficient matrix\cite{Suzuki20111904,Suzuki20121074,Suzuki2013314,Suzuki20134279}.
Note that this symmetry yields Onsager reciprocal relations\cite{PhysRev.37.405}.
Generally, we can give $\Theta$ by 
\begin{equation}
\Theta(\boldsymbol{\tilde{u}})=\int_{0}^{\tilde{\boldsymbol{u}}}\boldsymbol{w}(\tilde{\boldsymbol{u}}')\cdot d\tilde{\boldsymbol{u}}'>0,\label{eq:suzuki}
\end{equation}
where $\boldsymbol{w}$ is a function of $\tilde{\boldsymbol{u}}'$,
because there is no entropy production in a equilibrium state, $\tilde{\boldsymbol{u}}=0$.

\subsection{A simple dissipative system\label{sub:A-damped-oscillator}}

\begin{figure}
\includegraphics[scale=0.5]{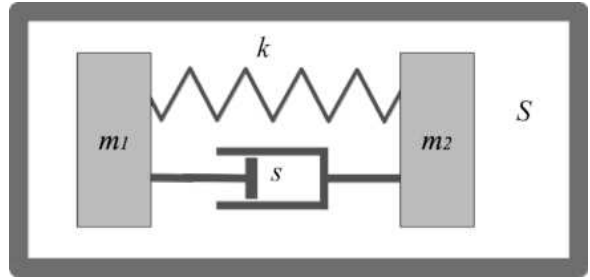}

\protect\caption{A simple example of dissipative systems.}
\label{fig:zu}
\end{figure}

Let us consider a damped coupled oscillators in a heat bath as an
example of the dissipative system in Sec.~\ref{sub:Second-law-of}.
The two oscillators are coupled via a purely viscous damper and purely
elastic spring connected in parallel as shown in Fig.~\ref{fig:zu}.
We write $q_{1}$ and $q_{2}$ for the position of each oscillator,
and $s$ for the entropy of dashpot. We give the Lagrangian (\ref{eq:tildL-3})
by
\begin{equation}
\!\frac{m_{i}}{2}u_{i}^{2}\!-\!\left(\!\frac{k}{2}(q_{1}\!-\! q_{2})^{2}\!+\!\epsilon(s)\!+\! E(S)\!\right)\!+\! p_{i}\!\left(\!\frac{dq_{i}}{dt}\!-\! u_{i}\!\right)\!\!,\label{eq:Lagrangian-3-1}
\end{equation}
where $m_{i}$ is the mass of each oscillator, $k$ is the spring
constant, and $\epsilon$ and $E$ are respectively the internal energies
of the dashpot and the environment. The constraints of entropy are
given by Eqs.~(\ref{eq:nonholo}) and (\ref{eq:nonholo-2}). This
system has the translation symmetries in space and time. From Eq.~(\ref{eq:ff}),
$f_{i}$ is given by $f_{1}=-f_{2}$ . Ignoring the cross terms between
$\boldsymbol{u}$ and $\left(1/T_{E}-1/T\right)$ in Eq.~(\ref{eq:quadro}),
we find that $f_{i}$ is the frictional forces in dashpot exerted
on the each oscillator, $f_{1}=-f_{2}=-\xi(u_{1}-u_{2})$ where $\xi$
is positive, and heat flow $Q$ satisfies
\begin{equation}
Q=\lambda\left(\frac{1}{T_{E}}-\frac{1}{T}\right),\label{eq:heat-flow}
\end{equation}
where $\lambda>0$ is thermal coefficient. The equation~(\ref{eq:heat-flow})
shows that heat flows from high temperature to low temperature. In
this way, we can determine the coefficients of the constraints of
entropy, (\ref{eq:nonholo}) and (\ref{eq:nonholo-2}) by the symmetries
and the second law of thermodynamics without knowing the exact form
of Lagrangian.

In addition, the Lagrangian (\ref{eq:Lagrangian-3-1}) also satisfies
Galilean symmetry if $m_{i}$ is constant. The Galilean transformation
is given by $\boldsymbol{g_{q}}=t$, $g_{p_{i}}=m_{i}$, and $g_{s}=g_{t}=0$.
The moment map $G=\sum_{i}(tp_{i}-m_{i}q_{i})$ is conserved.

\section{Applications to fluids\label{sec:Advanced-theory-for}}

We apply the theory in the previous section to the realistic fluids
in an inhomogeneous temperature field. We focus on the equation of
entropy,
\begin{equation}
\frac{\partial}{\partial t}(\rho s)=\Theta-\nabla\cdot\boldsymbol{J}.\label{eq:entropylaw}
\end{equation}
Here $\rho$, $s$, $\Theta$, and $\boldsymbol{J}$ denote mass density,
specific entropy, dissipative function, and entropy flux, respectively.
In our previous theory\cite{Fukagawa01052012}, we showed that the
equations of motion can be derived from actions with the aid of Eq.~(\ref{eq:entropylaw}).
In this section, we use the symmetries and the second law of thermodynamics
to obtain the dissipative function $\Theta$, and moreover we determine
entropy flux $\boldsymbol{J}$ by considering the well-posedness of
physical system.

First we show the simple example of deriving Eq.~(\ref{eq:entropylaw})
in a Newtonian fluid by using symmetries and the second law in Sec.~\ref{sec:Newtonian-fluid}.
Next, we discuss the fluids with interfaces in Sec.~\ref{sec:Interface-energy}.
We give the details in the case of a one-component fluid, a two-component
fluid, and a chiral liquid crystal.

\subsection{Newtonian fluid\label{sec:Newtonian-fluid}}

We write $\boldsymbol{A}=(A_{1},A_{2},A_{3})$ for the initial position
of a fluid particle. By the definition, the material derivative $D_{t}\equiv\partial_{t}+\boldsymbol{v}\cdot\nabla\ $
of $A_{i}$ is zero:
\begin{equation}
\frac{\partial A_{i}}{\partial t}+\boldsymbol{v}\cdot\nabla A_{i}=0.\label{eq:Clebsch}
\end{equation}
Here and below, Roman indexes run from 1 to 3. In terms of the control
theory, Eq.~(\ref{eq:Clebsch}) shows that the velocity field $\boldsymbol{v}$
controls the state $\boldsymbol{A}$. Mass density obeys the mass
conservation law:
\begin{equation}
\rho(t,\boldsymbol{x})-\rho_{{\rm init}}J^{-1}=0,\label{eq:massCon}
\end{equation}
where $\rho_{{\rm init}}$ is the initial mass density and $J^{-1}\equiv\partial(A_{1},A_{2},A_{3})/\partial(x_{1},x_{2},x_{3})\neq0$
denotes the contraction percentage of a fluid particle. This conservation
law (\ref{eq:massCon}) is required for the Lagrangian of Newtonian
fluid to be invariant under Galilean transformation\cite{Kambe200798}
as discussed in Sec.~\ref{sub:A-damped-oscillator}. The time derivative
of Eq.~(\ref{eq:massCon}) yields
\begin{equation}
\frac{\partial\rho}{\partial t}+\nabla\cdot(\rho\boldsymbol{v})=0,\label{eq:conservation law of rho}
\end{equation}
with the aid of Eq.~(\ref{eq:Clebsch}). See the details in App.~\ref{sec:Masscon}.
We assume the local thermal equilibration, and write $s$ for specific
entropy, i.e., entropy density per unit of mass. Then, the dynamics
is described by the trajectory in the space $(\boldsymbol{A}(\boldsymbol{x}),s(\boldsymbol{x}),t)$.
The nonholonomic constraint of entropy is given in the differential
form of $ds,dA_{i},\partial_{i}dA_{j},$ and $dt$. Here we write
$\partial_{j}$ for $\partial/\partial x_{j}$. Let us give it as
\begin{equation}
T\rho ds+M_{i}dX_{i}+\sigma_{ij}\partial_{i}(dX_{j})+Qdt=0.\label{eq:deltaS}
\end{equation}
Here $dX_{j}$ and $\partial_{i}(dX_{j})$ are short-hand notations
for 
\begin{equation}
dX_{j}\!=\!-\frac{\partial x_{j}}{\partial A_{i}}dA_{i}\ {\rm and}\ \partial_{i}dX_{j}\!=\!-\partial_{i}\left(\!\frac{\partial x_{j}}{\partial A_{k}}dA_{k}\!\right),\label{eq:deltaX}
\end{equation}
respectively. The higher differentiations of $\boldsymbol{X}$ are
not required in the Newtonian fluid. If we consider the fluids with
interfaces, these higher differentiations are required as we will
explain in Sec.~\ref{sec:Interface-energy}. The coefficients $M_{j}$,
$\sigma_{ij}$, and $Q$ are determined to be consist with symmetries
and the second law of thermodynamics. Space translation symmetry is
given by $(g_{s},g_{A_{i}},g_{t})=(\partial_{j}s,\partial_{j}A_{i},0)$,
i.e., 
\begin{equation}
\phi'(\boldsymbol{x}',t)=\phi(\boldsymbol{x},t)\ {\rm where}\ \boldsymbol{x}'=\boldsymbol{x}-\boldsymbol{\alpha}\label{eq:e}
\end{equation}
and $\phi$ denotes $s$ and $A_{i}$. It turns Eq.~(\ref{eq:deltaS})
into
\begin{equation}
T\rho\partial_{j}s-M_{j}=0,\label{eq:deltaS-6}
\end{equation}
with the aid of Eq.~(\ref{eq:deltaX}). The infinitesimal rotational
translation is given by $(g_{s},g_{A_{i}},g_{t})=(\epsilon_{jkl}x_{k}\partial_{l}s,\epsilon_{jkl}x_{k}\partial_{l}A_{i},0)$,
i.e., 
\begin{equation}
\phi'(\boldsymbol{x}',t)=\phi(\boldsymbol{x},t),\ {\rm where}\ \boldsymbol{x}'=\boldsymbol{x}-\boldsymbol{\alpha}\times\boldsymbol{x}.\label{eq:rot0}
\end{equation}
Here, $\epsilon_{ijk}$ is the Levi-Civita symbol. It proves $\sigma$
to be a symmetric tensor $\sigma^{S}$:
\begin{equation}
\epsilon_{ijk}\sigma_{jk}=0,\label{eq:asym}
\end{equation}
with the aid of Eqs.~(\ref{eq:deltaS}) and (\ref{eq:deltaX}). Time
translation, $(g_{s},g_{A_{i}},g_{t})=(\partial_{t}s,\partial_{t}A_{i},1)$
yields 
\begin{equation}
\frac{\partial}{\partial t}(\rho s)=\frac{1}{T}\sigma_{ij}^{S}e_{ij}-\nabla\cdot(\rho s\boldsymbol{v})-\frac{Q}{T},\label{eq:bs-2}
\end{equation}
with the aid of Eqs.~(\ref{eq:Clebsch}), (\ref{eq:conservation law of rho}),
(\ref{eq:deltaS}), (\ref{eq:deltaS-6}), and (\ref{eq:asym}). Here
$e_{ij}$ denotes the strain rate tensor:
\begin{equation}
e_{ij}\equiv\frac{1}{2}\left(\partial_{j}v_{i}+\partial_{i}v_{j}\right).\label{eq:eij}
\end{equation}
Using the same procedure in Eq.~(\ref{eq:enecon}), we have
\begin{equation}
\frac{\partial}{\partial t}{\cal H}+\partial_{j}\left[\left\{ ({\cal H}+P)\delta_{ij}+\sigma_{ij}^{S}\right\} v_{i}\right]+Q=0.\label{eq:EnergyCON}
\end{equation}
The Hamiltonian ${\cal H}$ and $P$ denote the total energy density
and pressure, respectively. See the details in App.~\ref{sec:The-Navier-Stokes}.
To satisfies energy conservation law, $Q$ should be given in this
form,
\begin{equation}
Q=\nabla\cdot\boldsymbol{J}_{q},\label{eq:NJ}
\end{equation}
where $\boldsymbol{J}_{q}$ is considered as heat flux. We can rewrite
Eq.~(\ref{eq:bs-2}) in the form of Eq.~(\ref{eq:entropylaw}).
The dissipative function $\Theta$ and entropy flux $\boldsymbol{J}$
are respectively given by
\begin{eqnarray}
\Theta & = & \frac{1}{T}\sigma_{ij}^{S}e_{ij}+\boldsymbol{J}_{q}\cdot\nabla\left(\frac{1}{T}\right)>0,\label{eq:dissipative}\\
\boldsymbol{J} & = & \rho s\boldsymbol{v}+\frac{\boldsymbol{J}_{q}}{T}.\label{eq:EntropyFlux}
\end{eqnarray}
The dissipative function $\Theta$ denotes the entropy production
rate, and should be positive because of the second law. Considering
$\Theta$ is zero in a equilibrium state, $\Theta$ is written as
Eq.~(\ref{eq:suzuki}) where $\tilde{\boldsymbol{u}}\equiv(\boldsymbol{v},\nabla T)$.
Low approximation of $\Theta$ is given in the quadric form, and thus
$\sigma$ and $\boldsymbol{J}_{q}$ are given by linear combination
of $e_{ij}$ and $\nabla T$,
\begin{eqnarray}
\sigma_{ij}^{S} & = & 2\zeta_{{\rm s}}e_{ij}+(\zeta_{{\rm b}}-2\zeta_{{\rm s}}/3)\delta_{ij}e_{kk},\label{eq:sigma}\\
\boldsymbol{J}_{q} & = & -\lambda\nabla T,\label{eq:heat}
\end{eqnarray}
where $\zeta_{{\rm s}}$, $\zeta_{{\rm b}}$ and $\lambda$ are the
coefficients of shear and bulk viscosities, and thermal conductivity,
respectively. Here, we ignore the cross terms between $e_{ij}$ and
$\nabla T$. All the coefficients $\zeta_{{\rm s}}$ , $\zeta_{{\rm b}}$
and $\lambda$ are positive because of Eq.~(\ref{eq:dissipative}).

Finally we obtain the equation of entropy (\ref{eq:entropylaw}) from
the symmetries, and the second law without knowing the exact Lagrangian.
The equation of motion can be derived by minimizing an action with
subject to the constraint of entropy. See the details in App.~\ref{sec:The-Navier-Stokes}.

\subsection{Interface energy\label{sec:Interface-energy}}

The new point in this section is the determination of the entropy
flux $\boldsymbol{J}$ related to interface energy. We show the details
in the case of vaporization, dissolution, and the rotation of a chiral
liquid crystals.

\subsubsection{Vaporization in a one-component fluid\label{sec:GAS-LIQUID}}

The mass density $\rho$ jumps from high to low at liquid-gas interface.
Thus the interface energy is defined as an excessive internal energy
per unit area induced by the density gradient in the interface zone\cite{JDvanderWaals}.
The total internal energy is given by the sum of the bulk internal
energy $\rho\epsilon(\rho,s)$ and the interface internal energy $E(\rho,\nabla\rho)$:
\begin{equation}
\rho\epsilon+E.\label{eq:newene}
\end{equation}
The interface energy $E$ is determined by the intermolecular forces,
and then scarcely depends on the entropy\cite{israelachvili2011intermolecular}.
The Lagrangian density is given by
\begin{equation}
{\cal L}\equiv\rho\left(\frac{1}{2}\boldsymbol{v}^{2}-\epsilon\right)-E.\label{eq:Le-1}
\end{equation}
The action is given by the integral of Eq.~(\ref{eq:Le-1}) over
the considered space and time. The equations of motion is derived
from the necessary condition of minimizing the action with subject
to the constraints (\ref{eq:Clebsch}), (\ref{eq:massCon}) and the
constraint for entropy. The derivative of the interface internal energy
$E(\rho,\nabla\rho)$ yields
\begin{eqnarray}
 & \!\! & \frac{\partial E}{\partial\rho}d\rho+\frac{\partial E}{\partial\nabla\rho}\cdot d(\nabla\rho)\nonumber \\
 & \!=\! & \!\left\{ \!\frac{\partial E}{\partial\rho}\!-\!\nabla\!\cdot\!\left(\frac{\partial E}{\partial\nabla\rho}\right)\!\right\} \! d\rho\!+\!\nabla\!\cdot\left(\!\frac{\partial E}{\partial\nabla\rho}d\rho\!\right)\!.\label{eq:dE}
\end{eqnarray}
The last term in the second line of Eq.~(\ref{eq:dE}),
\begin{equation}
\nabla\cdot\left(\frac{\partial E}{\partial\nabla\rho}d\rho\right),\label{eq:surface}
\end{equation}
does not vanish by itself. We cannot impose any boundary condition
on the mass density $\rho$ to cancel Eq.~(\ref{eq:surface}), because
the density only satisfy a pure transport Eq.~(\ref{eq:conservation law of rho}).
If we impose an excessive boundary condition artificially, it will
violate the well-posedness of the system, i.e., the existence of the
solution in this case. Note that the terminology ``boundary condition''
does not mean the liquid-gas interface. As we mentioned at the beginning
of this section, the liquid-gas interface are expressed by the area
where the mass density $\rho$ jumps from high to low.

Without loss of generality, we can give the constraint of entropy
in the form of
\begin{equation}
T\rho ds+M_{i}dX_{i}+\sigma_{ij}\partial_{i}(dX_{j})+Qdt+R=0.\label{eq:deltaS-1}
\end{equation}
The coefficients, $M_{i}$, $\sigma_{ij}$, and $Q$ are respectively
given by Eqs.~(\ref{eq:deltaS-6}), (\ref{eq:asym}), and (\ref{eq:NJ})
for the consistency of the symmetries. The new one-form $R$ is determined
so as to cancel Eq.~(\ref{eq:surface}) in the calculus of the necessary
condition of minimizing action without breaking the symmetries. With
the aid of the second law, Eqs.~(\ref{eq:deltaS-1}) yields the equation
of entropy in the form of Eq.~(\ref{eq:entropylaw}). Using the same
procedure, the dissipative function $\Theta$ is uniquely given by
Eq.~(\ref{eq:dissipative}) by considering $\Theta$ is a function
of $\boldsymbol{\tilde{u}}=(\boldsymbol{v},\nabla T)$. Thus the one-form
$R$ has to relate only to the entropy flux, 
\begin{equation}
\boldsymbol{J}=\rho s\boldsymbol{v}+\frac{\boldsymbol{J}_{q}}{T}+\boldsymbol{J}_{s}.\label{eq:entropyflux-1}
\end{equation}
Here, $\boldsymbol{J}_{s}$ is the entropy flux derived from $R$.
We find that if $\boldsymbol{J}_{s}$ is given by 
\begin{equation}
\boldsymbol{J}_{s}=\frac{1}{T}\frac{\partial E}{\partial\nabla\rho}D_{t}\rho,\label{eq:Js}
\end{equation}
the surface term (\ref{eq:surface}) is canceled in the variational
calculus without breaking the symmetries. Let us make sure of it.
Replacing the time derivative $\partial_{t}$ by $d$ of $T\nabla\cdot\boldsymbol{J}_{s}$
whose $\boldsymbol{J}_{s}$ is Eq.~(\ref{eq:Js}) yields
\begin{equation}
R\equiv T\nabla\cdot\left\{ \frac{1}{T}\frac{\partial E}{\partial\nabla\rho}\left(d\rho+(\nabla\rho)\cdot d\boldsymbol{X}\right)\right\} ,\label{eq:dJs}
\end{equation}
with the aid of Eq.~(\ref{eq:Clebsch}). The term $T\nabla\cdot\left\{ \frac{1}{T}\frac{\partial E}{\partial\nabla\rho}d\rho\right\} $
in Eq.~(\ref{eq:dJs}) cancels the surface term (\ref{eq:surface}).
Next, replacing $d$ by $\partial_{i}$ make it vanish because this
replacement of $d\rho+(\partial_{j}\rho)dX_{j}$ yields $(\partial_{i}-\delta_{ij}\partial_{j})\rho=0$
with the aid of Eq.~(\ref{eq:deltaX}). Thus Eq.(\ref{eq:Js-1})
has the space translation symmetry. Similarly, it also has the rotational
symmetry too. Thus we find out that Eq.~(\ref{eq:Js}) is appropriate
for the Lagrangian (\ref{eq:Le-1}), and obtain the equation of motion,
\begin{equation}
\frac{\partial}{\partial t}\left(\rho v_{i}\right)+\partial_{j}\left(\rho v_{i}v_{j}+\Pi_{ij}+\sigma{}_{ij}^{S}\right)-\gamma_{i}=0.\label{eq:momentum}
\end{equation}
Here we write $\gamma_{i}$ and $\Pi_{ij}$ respectively for
\begin{eqnarray}
\gamma_{i} & \equiv & \frac{\partial_{j}T}{T}\left(\frac{\partial E}{\partial\partial_{i}\rho}\partial_{j}\rho-\frac{\partial E}{\partial\partial_{j}\rho}\partial_{i}\rho\right),\label{eq:gamma}\\
\Pi_{ij} & \equiv & \left\{ P-\rho T\nabla\cdot\left(\frac{1}{T}\frac{\partial E}{\partial\nabla\rho}\right)+\rho\frac{\partial E}{\partial\rho}-E\right\} \delta_{ij}\nonumber \\
 &  & +\frac{\partial E}{\partial\partial_{i}\rho}\partial_{j}\rho.\label{eq:PI}
\end{eqnarray}
Note that $\boldsymbol{\gamma}$ vanishes on the condition of satisfying
the chiral (mirror) symmetry, 
\begin{equation}
\frac{\partial E}{\partial\partial_{j}\rho}\partial_{k}\rho=\frac{\partial E}{\partial\partial_{k}\rho}\partial_{j}\rho.\label{eq:iso}
\end{equation}
See the details in App.~\ref{sec:GAS-LIQUID-1}. If we define the
interface internal energy as $E=(K\left|\nabla\rho\right|^{2})/2$,
where $K$ is a function of $\rho$, Eq.~(\ref{eq:Js}) is equivalent
to the theory which Onuki named ``The Dynamic van der Waals model''\cite{0295-5075-84-3-36003,PhysRevE.75.03630s4}.
The entropy flux (\ref{eq:Js}) caused along with the moving interface
is related to the latent heat transfer across the liquid-gas interface.
For example, suppose that a drop of water shrinks by vaporization,
and then its surface area also decreases. The entropy of liquid area
transfers to gas area across the interface. This motion involves the
energy transfer, because the increasing and decreasing of entropy
causes the ones of the internal energy.

In the microscopic view, the high energy liquid molecules can go into
the gas phase only through the surface of the drop, and take energy
out of the surrounding molecules. Thus vaporization takes latent heat,
i.e., entropy from the liquid phase to the gas phase. The energy required
to free a molecule from the liquid is equivalent to the energy needed
to break the molecular bonds and keep molecules away from surface.
Therefore, the intermolecular forces that determine the interface
internal energy of a substance are the same as those that determine
its latent heat and boiling point\cite{israelachvili2011intermolecular}.

\subsubsection{Dissolution in a two-component fluid\label{sec:Two-component-Fluid}}

We consider a two-component fluid composed of two substances: solute
and solvent. The mass densities $\rho_{{\rm solute}}$ and $\rho_{{\rm solvent}}$
are changed by the mass average velocity $\boldsymbol{v}$, and the
diffusion flux $\boldsymbol{j}_{c}$ , as
\begin{eqnarray}
\partial_{t}\rho_{c}+\nabla\cdot(\rho_{c}\boldsymbol{v}+\boldsymbol{j}_{c}) & = & 0,\label{eq:kakusannryu}
\end{eqnarray}
where $c$ denotes solute or solvent. The diffusion flux $\boldsymbol{j}_{c}$
denotes the amount of the component transported by diffusion through
unit area in unit time. Note that the diffusion flux $\boldsymbol{j}_{c}$
satisfies $\boldsymbol{j}_{{\rm solute}}=-\boldsymbol{j}_{{\rm solvent}}$
because the sum of each mass conservation law (\ref{eq:kakusannryu})
yields Eq.~(\ref{eq:conservation law of rho}), where $\rho\equiv\rho_{{\rm solute}}+\rho_{{\rm solvent}}$
is the total mass density. Hereafter we write $\boldsymbol{j}$ for
$\boldsymbol{j}_{{\rm solute}}$. The diffusion flux $\boldsymbol{j}$
describes the relative motion of the solute and the solvent. The conservation
laws (\ref{eq:kakusannryu}) are rewritten into the set of Eq.~(\ref{eq:massCon})
and the diffusion equation of the solute,
\begin{equation}
\rho D_{t}\psi+\nabla\cdot\boldsymbol{j}=0,\label{eq:diffusion}
\end{equation}
where $\psi\equiv\rho_{a}/\rho$ is the mass fraction of the solute.
Let $\boldsymbol{a}$ be the amount of the solute flowing through
the unit interface of a fluid particle, i.e., $D_{t}a_{i}-j_{j}J_{i}^{j}=0$,
where $J_{i}^{j}$ is the cofactor of $\partial x_{j}/\partial A_{i}$.
We can rewrite it into 
\begin{equation}
(D_{t}a_{i})J^{-1}-\boldsymbol{j}\cdot\nabla A_{i}=0.\label{eq:Dta}
\end{equation}
We also fix the value of $\boldsymbol{a}$ at the boundary. The derivative
of the interface energy density $E(\rho,\psi,\nabla\psi)$ yields

\begin{equation}
\nabla\cdot\left(\frac{\partial E}{\partial\nabla\psi}d\psi\right),\label{eq:surface-1}
\end{equation}
with the aid of integration by parts as same as Eq.~(\ref{eq:surface}).
The constraint of entropy is given in the form of
\begin{equation}
T\rho ds+M_{i}dX_{i}+\sigma_{ij}\partial_{i}(dX_{j})+\nu_{i}da_{i}+Qdt+R=0.\label{eq:deltaS-1-1}
\end{equation}
Using the same procedure in Sec.~\ref{sec:GAS-LIQUID}, we obtain
$M_{i}=T\rho\partial_{i}s+\nu_{j}\partial_{i}a_{j}$ by the space
translate symmetry, and Eqs.~(\ref{eq:asym}) and (\ref{eq:NJ})
for $\sigma_{ij}$, and $Q$ by the rotational symmetry and the time
translate symmetry, respectively. By the second law, Eq.~(\ref{eq:deltaS-1-1})
can be rewritten in the form of Eq.~(\ref{eq:entropylaw}). The dissipative
function $\Theta$ is given by
\begin{equation}
\Theta=\frac{\sigma_{ij}^{S}e_{ij}+\tilde{\boldsymbol{\nu}}\cdot\boldsymbol{j}}{T}+\boldsymbol{J}_{q}\cdot\nabla\left(\frac{1}{T}\right).\label{eq:theta}
\end{equation}
Here, we use Eq.~(\ref{eq:Dta}) and define $\tilde{\boldsymbol{\nu}}$
as $\tilde{\boldsymbol{\nu}}\equiv J\nu_{i}\nabla A_{i}$. In the
low degree approximation, Eq.~(\ref{eq:theta}) is given by the quadratic
form of $e_{ij}$, $\boldsymbol{j}$, and $\nabla T$. Without loss
of generality, we have
\begin{eqnarray}
\tilde{\boldsymbol{\nu}} & \!\!\!\!=\!\!\!\! & \xi\boldsymbol{j}+\eta\nabla T,\label{eq:nu}\\
\boldsymbol{J_{q}} & \!\!\!\!=\!\!\!\! & -\eta T\boldsymbol{j}-\lambda\nabla T,\label{eq:HeatFlux}
\end{eqnarray}
where $\xi$ is the coefficient of friction for the diffusion flux
$\boldsymbol{j}$, and $\lambda$ is the coefficient of thermal conductivity
as in $\S$ 58 of \cite{landau1959fm}. The coefficient $\eta$ in
Eq.~(\ref{eq:nu}) expresses the Soret effect describing the flow
of the solute induced by a temperature gradient. On the other hand,
the coefficient $\eta$ in Eq.~(\ref{eq:HeatFlux}) shows the Dufour
effect describing the energy flux due to the diffusion flux $\boldsymbol{j}$
occurring\cite{PhysRev.37.405}. On the other hand, the entropy flux
$\boldsymbol{J}$ is given by Eq.~(\ref{eq:entropyflux-1}), where
\begin{equation}
\boldsymbol{J}_{s}=\frac{1}{T}\frac{\partial E}{\partial\nabla\psi}D_{t}\psi\label{eq:Js-1}
\end{equation}
is determined to cancel Eq.~(\ref{eq:surface-1}). Because $\partial E/\partial\nabla\psi$
takes large absolute value at the interface, and $D_{t}\psi$ expresses
the moving of the interface, Eq.~(\ref{eq:Js-1}) shows that entropy
flux occurs with accompanying the moving interface, which is related
to the heat of dissolution\cite{israelachvili2011intermolecular}.

Then, by using the variational principle with the constraint of entropy,
we obtain the equations of motion for the mass average velocity $\boldsymbol{v}$,
and the diffusion flux $\boldsymbol{j}$. The former is
\begin{equation}
\frac{\partial}{\partial t}\left(\rho v_{i}\right)+\partial_{j}\left(\rho v_{i}v_{j}+\Pi_{ij}+\sigma_{ij}^{S}\right)+\gamma_{i}=0,\label{eq:momentum-2-1}
\end{equation}
where we write $\gamma_{i}$ and $\Pi_{ij}$ for
\begin{eqnarray}
\gamma_{i} & = & \frac{\partial_{j}T}{T}\left(\frac{\partial E}{\partial\partial_{i}\psi}\partial_{j}\psi-\frac{\partial E}{\partial\partial_{j}\psi}\partial_{i}\psi\right),\label{eq:gamma-1}\\
\Pi_{ij} & = & \left(P+\rho\frac{\partial E}{\partial\rho}-E\right)\delta_{ij}+\frac{\partial E}{\partial\partial_{i}\psi}\partial_{j}\psi.
\end{eqnarray}
The equation (\ref{eq:gamma-1}) vanishes if the interface energy
$E$ has the chiral (mirror) symmetry, i.e.,
\begin{equation}
\frac{\partial E}{\partial\partial_{i}\psi}\partial_{j}\psi=\frac{\partial E}{\partial\partial_{j}\psi}\partial_{i}\psi.\label{eq:iso-1}
\end{equation}
as same as Eq.~(\ref{eq:iso}). The latter is
\begin{eqnarray}
\!\!\!\!\!\!\!\!\!\!\!\!\!\! & \!\!\!\! & \!\! L_{\boldsymbol{v}}^{1}\left\{ \frac{1}{\rho}\left(\frac{1}{\psi}+\frac{1}{1-\psi}\right)\boldsymbol{j}\right\} \nonumber \\
\!\!\!\!\!\!\!\!\!\!\!\!\!\! & \!\!=\!\! & \!\!-\nabla\left\{ \mu^{*}+\frac{1}{2\rho^{2}}\left(\frac{1}{\psi^{2}}-\frac{1}{(1-\psi)^{2}}\right)\boldsymbol{j}^{2}\right\} -\boldsymbol{\tilde{\nu}},\label{eq:DiffusionFlux-1}
\end{eqnarray}
where $L_{\boldsymbol{v}}^{1}\equiv\partial_{t}+\nabla(\boldsymbol{v}\cdot\ )-\boldsymbol{v}\times\nabla\times\ $
denotes the convected time derivative for the cotangent vector, and
considered as the Lie derivative in mathematics\cite{bernard1984geometrical},
and $\mu^{*}$ is the generalized chemical potential defined as
\begin{equation}
\mu^{*}\equiv\mu+\frac{1}{\rho}\frac{\partial E}{\partial\psi}-\frac{T}{\rho}\nabla\cdot\left(\frac{1}{T}\frac{\partial E}{\partial\nabla\psi}\right).\label{eq:GeneraizedMu}
\end{equation}
See the details in App.~\ref{sec:Two-component-Fluid-1}. Our variational
principle clarifies that the entropy flux (\ref{eq:Js-1}) caused
by the dissolution yields an inhomogeneous temperature field, and
it effects on the dissolution via the generalized chemical potential
(\ref{eq:GeneraizedMu}). If we ignore an inhomogeneous temperature
and heat flux, the equations (\ref{eq:momentum-2-1}) and (\ref{eq:DiffusionFlux-1})
are the same of the theories based on free energies\cite{JDvanderWaals,cahn1958free}.

\subsubsection{Rotation of a chiral liquid crystal\label{sec:LC}}

We write $\boldsymbol{n}$ for the director of a liquid crystals,
which is a dimensionless unit vector. The angular velocity $\tilde{\boldsymbol{\omega}}$
denotes the rotation of the director. The material time derivative
of $\boldsymbol{n}$ is given by
\begin{equation}
D_{t}n_{i}=\omega_{i},\label{eq:omega}
\end{equation}
where $\boldsymbol{\omega}$ is defined as $\boldsymbol{\omega}\equiv\tilde{\boldsymbol{\omega}}\times\boldsymbol{n}$.
The equation above expresses a kind of the rigid motion because the
norm of $\boldsymbol{n}$ is constant, $|\boldsymbol{n}|=1$. To simplify
the discussion, we neglect the surface energy at first. We modify
the nonholonomic constraint (\ref{eq:deltaS}) as
\begin{equation}
T\rho ds+M_{j}dX_{j}+\sigma_{ij}(d\partial_{i}X_{j})+g_{i}dn_{i}+Qdt=0.\label{eq:deltaS-3}
\end{equation}
where the term $g_{i}dn_{i}$ is added for the dependency of $\boldsymbol{n}$.
We have $M_{j}=T\rho\partial_{j}s+g_{i}\partial_{j}n_{i}$ because
of the space translation symmetry as in Eq.~(\ref{eq:deltaS-6}),
and the coefficient $Q$ is given by Eq.~(\ref{eq:NJ}) because of
the energy conservation law. We divided $\sigma$ into symmetric part
$\sigma^{S}$ and asymmetrical part $\sigma^{A}$. The rotational
symmetry yields
\begin{equation}
\epsilon_{ijk}(\sigma_{jk}^{A}+g_{j}n_{k})=0.\label{eq:asym-1}
\end{equation}
Here we use Eq.~(\ref{eq:rot0}) and the rotational transformation
for vector,
\begin{equation}
n_{\alpha}'(\boldsymbol{x}')=n_{\alpha}(\boldsymbol{x})-\boldsymbol{\alpha}\times\boldsymbol{n}\ {\rm where}\ \boldsymbol{x}'=\boldsymbol{x}-\boldsymbol{\alpha}\times\boldsymbol{x}.\label{eq:rot-1}
\end{equation}
Using the same procedure in Sec.~\ref{sec:GAS-LIQUID}, we can rewrite
Eq.~(\ref{eq:deltaS-3}) in the form of Eq.~(\ref{eq:entropylaw}).
The entropy flux $\boldsymbol{J}$ and the dissipative function $\Theta$
are respectively given by Eq.~(\ref{eq:EntropyFlux}) and
\begin{eqnarray}
\Theta & = & \frac{1}{T}(\sigma_{ij}^{S}e_{ij}+\sigma_{ij}^{A}d_{ij}+g_{i}\omega_{i})+\boldsymbol{J_{q}}\cdot\nabla\left(\frac{1}{T}\right)\nonumber \\
 & = & \frac{1}{T}(\sigma_{ij}^{S}e_{ij}+g_{i}N_{i})+\boldsymbol{J_{q}}\cdot\nabla\left(\frac{1}{T}\right)>0\label{eq:dissipation-1}
\end{eqnarray}
where $e_{ij}$, $d_{ij}$ and $N_{i}$ are respectively defined as
Eq.~(\ref{eq:eij}),
\begin{equation}
d_{ij}\equiv\frac{1}{2}\left(\partial_{j}v_{i}-\partial_{i}v_{j}\right),\label{eq:dij}
\end{equation}
and 
\begin{equation}
N_{i}\equiv\omega_{i}-d_{ij}n_{j}.\label{eq:NW}
\end{equation}
We use Eq.~(\ref{eq:asym-1}) at the second line in Eq.~(\ref{eq:dissipation-1}).
The equation~(\ref{eq:NW}) expresses the rotation of a director
except the rotation of velocity. We have
\begin{eqnarray}
\boldsymbol{g} & = & \xi\boldsymbol{N}+\eta\nabla T,\label{eq:nu-1}\\
\boldsymbol{J_{q}} & = & \eta T\boldsymbol{N}-\lambda\nabla T,\label{eq:HeatFlux-1}
\end{eqnarray}
in the low degree approximation as discussed in Eqs.~(\ref{eq:nu})
and (\ref{eq:HeatFlux}). Here, $\xi$, $\eta$, and $\lambda$ are
rotational viscosity, Lehman coefficient, and heat conductivity. Next,
let us consider the interface energy $E(\nabla\rho,n_{i},\partial_{j}n_{i})$.
Using the same procedure in Sec.~\ref{sec:GAS-LIQUID}, we obtain
the equation of velocity field,
\begin{equation}
\rho\left\{ \!\frac{\partial}{\partial t}\boldsymbol{v}\!+\!\frac{1}{2}\nabla\!\boldsymbol{v}^{2}\!-\!\boldsymbol{v}\!\times\!(\nabla\!\times\!\boldsymbol{v})\!\right\} +\nabla\!:\!(\hat{\sigma}^{S}\!+\!\Pi)\!-\!\boldsymbol{\gamma}=0,\label{eq:Vmotion}
\end{equation}
where $\hat{\sigma}$, $\boldsymbol{\gamma}$, and $\Pi$ are respectively
given by
\begin{equation}
\hat{\sigma}_{ij}^{S}\equiv\rho\pi_{ki}\partial_{j}n_{k}+\sigma_{ij}^{S},\label{eq:h}
\end{equation}
and Eqs.~(\ref{eq:gamma}) and (\ref{eq:PI}). See the details in
App.~\ref{sec:NLCcal}. Yoshioka et al. found the rotation of a chiral
LC caused by a concave-convex surface\cite{C4SM00670D}. This experiment
suggests that another type of heat-driven unidirectional motion occurs
through an essentially different mechanism from the Lehmann. This
phenomenon can be expressed as the following. Suppose the surface
 energy $E$ includes the term, which do not satisfy Eq.~(\ref{eq:iso}),
such as $(K_{2}+K_{24})\nabla\rho\cdot\{\boldsymbol{n}\cdot\nabla\boldsymbol{n}+\boldsymbol{n}\times(\nabla\times\boldsymbol{n})\}$
where $K_{2}$ and $K_{24}$ are respectively twist and saddle-splay
elastic constants\cite{:/content/aip/journal/pof1/9/6/10.1063/1.1761821}.
The equations~(\ref{eq:gamma}) does not vanish: $\gamma\neq0$,
and thus the gradient temperature $\nabla T$ on the surface yields
a rotation of the velocity field (\ref{eq:Vmotion}), while the Lehman
effect yields the one of the director.

\section{Summary\label{sec:Discussion}}

Our method is based on the two principles. The first one is the variational
principle, stating that the realized motion minimizing an action.
The second one is the second law of thermodynamics, i.e., the entropy
of isolated systems increase. The equations of motion for fluids can
be derived by the stationary condition of the action with subject
to the equation of entropy\cite{Fukagawa01052012}. We give the equation
of entropy by using only few requirements in our theory, including
symmetries and well-posedness, in addition to the second law. The
resultant equations derived from our method automatically satisfy
the conservation laws associated with the symmetries because of the
generalized Noether's theorem in Sec.~\ref{sub:Noether's-theorem-with}.
The variational formalization should have a unique weak solution within
a certain class of functions with boundary conditions, because of
the well-posedness for the physical systems. For the cases of the
fluids with interface, it requires to cancel all the surface terms
yielding the excessive boundary conditions as we discussed in Sec.~\ref{sec:Interface-energy}.
This clarifies that moving interfaces generate entropy fluxes, i.e.,
induce inhomogeneous temperature fields, and then the dynamics of
fluids are affected by them. In the case of vaporization and dissolution,
we clarify the reaction between the entropy flux and these phenomena
connected by interface energy. We also give a new equation of a chiral
liquid crystal explaining its rotation caused by heat, which is totally
different from the well-known Lehman effect. While the original Lehman
effect is about the rotation of the director, our new theory explains
the rotation of the velocity field originated from interface energy.
The novelty of this article lies in the self-consistent derivation
of the system for inhomogeneous materials which involve multiple physical
effects from multiple scales, taking into account of various competitions
and couplings.

\section*{Acknowledgments}

The work had started when Fukagawa visited Department of Mathematics
in Penn State University during the period from November 2013 to June
2014. H. Fukagawa and C. Liu want to acknowledge the partial support
of the National Science Foundation (USA) grants DMS-1109107, DMS-1216938
and DMS-1159937. H. Fukagawa and T. Tsuji gratefully acknowledge the
support of the International Institute for Carbon-Neutral Energy Research
(WPI-I2CNER), sponsored by the World Premier International Research
Center Initiative (WPI), MEXT, Japan, and appreciate the support of
the SATREPS project by JICA-JST. H. Fukagawa appreciates Youhei Fujitani
for advices, Billy D. Jones for valuable comments about the symmetries,
Shogo Tanimura, Shin-itiro Goto, and Jun Yoshioka for the discussions
about variational principle, Noether's theorem, and LCs, respectively.

\appendix

\section{Variational principle\label{sec:differential-form}}

We can rewrite the action (\ref{eq:Iqpu-3}) into $\int_{C}\Xi$,
where the integral curve and the integrand are respectively given
by $C\equiv[\boldsymbol{r}(t)|t_{{\rm init}}\leqq t\leqq t_{{\rm fin}}]$
and $\Xi(\boldsymbol{r},t)\equiv p_{i}dq_{i}-\tilde{H}(\boldsymbol{r})dt$.
Let $C$ and $C_{\alpha}$ be an optimized trajectory and another
trajectory, respectively, such that each endpoint is connected. Using
Stokes' theorem, we find that 
\begin{equation}
\oint_{C-C_{\alpha}}\!\!\!\Xi=\iint_{S_{\alpha}}d\Xi=0\label{eq:Stokes}
\end{equation}
is satisfied, where $S_{\alpha}$ is the integral area such that $\partial S_{\alpha}=C-C_{\alpha}$
and $d\Xi=dp\wedge dq-d\tilde{H}\wedge dt$. Let $\tilde{\boldsymbol{X}}_{\tilde{H}}=\boldsymbol{X}_{\tilde{H}}+\frac{\partial}{\partial t}\ $
be the vector field for $\frac{d\boldsymbol{r}(t)}{dt}$. Here, $\boldsymbol{X}_{\tilde{H}}$
denotes $\frac{dq}{dt}\frac{\partial}{\partial q}+\frac{dp}{dt}\frac{\partial}{\partial p}+\frac{ds}{dt}\frac{\partial}{\partial s}+\frac{du}{dt}\frac{\partial}{\partial u}$.
Provided $S_{\alpha}$ is small, Eq.~(\ref{eq:Stokes}) is equivalent
to

\begin{equation}
d\Xi(\tilde{\boldsymbol{X}}_{\tilde{H}},\tilde{\boldsymbol{Y}}_{\alpha})=0.\label{eq:dXI}
\end{equation}
for the corresponding vector field $\tilde{\boldsymbol{Y}}_{\alpha}$.
Thus the necessary condition of minimizing the action (\ref{eq:Iqpu-3})
is given by
\begin{eqnarray}
\boldsymbol{0} & \!\!\!\!=\!\!\!\! & d\Xi(\tilde{\boldsymbol{X}}_{\tilde{H}},\ )\nonumber \\
 & \!\!\!\!=\!\!\!\! & dp\wedge dq(\tilde{\boldsymbol{X}}_{\tilde{H}},\ )\!-\!(\iota_{\tilde{\boldsymbol{X}}_{\tilde{H}}}d\tilde{H})dt\!+\! d\tilde{H}\!\iota_{\tilde{\boldsymbol{X}}_{\tilde{H}}}dt\nonumber \\
 & \!\!\!\!=\!\!\!\! & dp\wedge dq(\boldsymbol{X}_{\tilde{H}},\ )\!-\!(\iota_{\boldsymbol{X}_{\tilde{H}}}d\tilde{H})dt\!+\! d\tilde{H}.\label{eq:SRT}
\end{eqnarray}
Here $d$ is the exterior derivative, and $\iota_{\boldsymbol{X}}$
denotes the interior product of $\boldsymbol{X}$. Thus if $\boldsymbol{X}_{\tilde{H}}$
satisfies 
\begin{eqnarray}
dp\wedge dq(\boldsymbol{X}_{\tilde{H}},\ )+d\tilde{H} & = & \boldsymbol{0},\label{eq:h1}\\
\iota_{\boldsymbol{X}_{\tilde{H}}}d\tilde{H} & = & 0,\label{eq:h2}
\end{eqnarray}
it holds Eq.~(\ref{eq:Stokes}). With the aid of Eq.~(\ref{eq:nonholo}),
we obtain Eq.~(\ref{eq:Iqpu-4}). If Eq.~(\ref{eq:h1}) is satisfied,
Eq.~(\ref{eq:h2}) is automatically valid: $\iota_{\boldsymbol{X}_{\tilde{H}}}d\tilde{H}=-dp\wedge dq(\boldsymbol{X}_{\tilde{H}},\boldsymbol{X}_{\tilde{H}})=0$.

\section{Noether's theorem\label{sec:differential-form-1}}

Consider the necessary condition of the Lie derivative ${\cal L}_{\tilde{\boldsymbol{X}}_{G}}$
of $d\Xi^{*}=dp\wedge dq-dH\wedge dt$ is zero:

\begin{equation}
\boldsymbol{0}={\cal L}_{\tilde{\boldsymbol{X}}_{G}}d\Xi^{*}=d\iota_{\tilde{\boldsymbol{X}}_{G}}d\Xi^{*}.\label{eq:SRT-1}
\end{equation}
Here $H$ is the Hamiltonian given by Eq.~(\ref{eq:Hamilton}), and
$\tilde{\boldsymbol{X}}_{G}\equiv\boldsymbol{X}_{G}+g_{t}\partial_{t}$
is the infinitesimal translation. The equation~(\ref{eq:SRT-1})
is satisfied if $\iota_{\tilde{\boldsymbol{X}}_{G}}d\Xi^{*}$ is a
exact form, 
\begin{equation}
\iota_{\tilde{\boldsymbol{X}}_{G}}d\Xi^{*}=\iota_{\boldsymbol{X}_{G}}(dp_{i}\wedge dq_{_{i}})-(\iota_{\boldsymbol{X}_{G}}dH)dt+(dH)g_{t}=-dG\label{eq:dG}
\end{equation}
where $G$ is a function sometimes called moment map. Thus we have
\begin{eqnarray}
\boldsymbol{0} & \!\!=\!\! & \left(\! g_{p_{i}}\!+\! g_{t}\!\frac{\partial H}{\partial q_{i}}\!\right)\! dq_{i}\!-\!\left(\! g_{q_{i}}\!-\! g_{t}\!\frac{\partial H}{\partial p_{i}}\!\right)\! dp_{i}\nonumber \\
 &  & +g_{t}\frac{\partial H}{\partial s}ds-(\boldsymbol{X}_{G}H)dt+dG,\label{eq:dG-3-3}
\end{eqnarray}
where $g_{q_{i}},g_{p_{i}}g_{s}$ are the components of $\boldsymbol{X}_{G}$.
With the aid of Eq.~(\ref{eq:nonholo}), we obtain Eq.~(\ref{eq:dG-3})
from Eq.~(\ref{eq:dG-3-3})

\section{Mass conservation law\label{sec:Masscon}}

Calculating by means of the cofactors yields $\frac{\partial J^{-1}}{\partial(A_{i}/x_{j})}=J^{-1}\frac{\partial x_{j}}{\partial A_{i}},$
while some algebra yields $\partial_{j}\left(J^{-1}\frac{\partial x_{j}}{\partial A_{i}}\right)=0.$
Then we have
\begin{equation}
dJ^{-1}=J^{-1}\frac{\partial x_{j}}{\partial A_{i}}\frac{\partial dA_{i}}{\partial x_{j}}=\partial_{j}\left(J^{-1}\frac{\partial x_{j}}{\partial A_{i}}dA_{i}\right).\label{eq:J2}
\end{equation}
Because the initial mass density $\rho_{{\rm init}}$ is the function
of $A_{i}$, then we have
\begin{equation}
d\rho_{{\rm init}}=\frac{\partial\rho_{{\rm init}}}{\partial A_{i}}dA_{i}=\frac{\partial\rho_{{\rm init}}}{\partial x_{j}}\frac{\partial x_{j}}{\partial A_{i}}dA_{i}.\label{eq:rhoinit}
\end{equation}
From Eqs.~(\ref{eq:J2}) and (\ref{eq:rhoinit}), we obtain 
\begin{equation}
d\!\left(\!\rho_{{\rm init}}J^{-1}\!\right)\!=\!\frac{\partial\rho_{{\rm init}}}{\partial A_{i}}dA_{i}J^{-1}\!\!+\!\rho_{{\rm init}}dJ^{-1}\!\!=\!\partial_{j}\!\left(\!\rho\frac{\partial x_{j}}{\partial A_{i}}dA_{i}\!\right)\!,\label{eq:lhsmasscon}
\end{equation}
and then we can rewrite (\ref{eq:massCon}) into
\begin{equation}
d\rho-\partial_{j}\left(\rho\frac{\partial x_{j}}{\partial A_{i}}dA_{i}\right)=0.\label{eq:mass_con-1}
\end{equation}
Replacing $d$ by $\partial/\partial t$, Eq.~(\ref{eq:mass_con-1})
yields the well known mass conservation law (\ref{eq:conservation law of rho})
with the aid of Eq.~(\ref{eq:Clebsch}).

\section{The Navier Stokes equations\label{sec:The-Navier-Stokes}}

The specific internal energy $\epsilon$ is given by the function
of the mass density $\rho$ and specific entropy $s$. Then we have
$d\epsilon=-Pd\rho^{-1}+Tds$. Pressure $P$ and temperature $T$
are respectively given by 
\begin{equation}
P\equiv-\left(\frac{\partial\epsilon}{\partial\rho^{-1}}\right)_{s}\!\!=\rho^{2}\left(\frac{\partial\epsilon}{\partial\rho}\right)_{s}\!\!\ {\rm and}\ T\equiv\left(\frac{\partial\epsilon}{\partial s}\right)_{\rho}\label{eq:PT}
\end{equation}
where the subscripts $_{s}$ and $_{\rho}$ indicate variables fixed
in the respective partial differentiations. The Lagrangian density
is given by the kinetic energy minus inertial energy:
\begin{equation}
{\cal L}(\rho,\boldsymbol{v},s)\equiv\rho\left\{ \frac{1}{2}\boldsymbol{v}^{2}-\epsilon(\rho,s)\right\} ,\label{eq:Le}
\end{equation}
where $\rho\boldsymbol{v}^{2}/2$ is the kinetic energy density. With
the aid of Eqs.~(\ref{eq:Clebsch}) and (\ref{eq:massCon}), the
modified Lagrangian is given by the integral over the considered space
$V$,
\begin{equation}
\tilde{L}\!\equiv\!\!\int_{V}\!\!\! d^{3}\!\boldsymbol{x}\!\left\{ \!{\cal L}\!+\! K\!\left(\rho\!-\!\rho_{{\rm init}}J^{-\!1}\right)\!+\!\beta_{i}\!\left(\frac{\partial}{\partial t}A_{i}\!+\!\boldsymbol{v}\cdot\!\nabla\! A_{i}\right)\!\right\} \!,\label{eq:action-4}
\end{equation}
with using undetermined multipliers, $K$ and $\beta_{i}$. The Lagrangian
(\ref{eq:action-4}) has the translation symmetries in time and space
because it does not depend on time and space explicitly, has rotational
symmetry because all the terms in Eq.~(\ref{eq:Le}) are scalars,
and has Galilean symmetry because of mass conservation law, similarly
in Sec.~\ref{sub:A-damped-oscillator}. The action is given by the
integral of Eq.~(\ref{eq:action-4}). Using Stokes' theorem (\ref{eq:Stokes}),
we have
\begin{eqnarray}
\!\!\!\!\!\!\!\!\!\! &  & \int_{V}\!\!\!\! dV\Biggl[\left(\rho\boldsymbol{v}\!+\!\beta_{j}\!\nabla\! A_{j}\right)\!\cdot\! d\boldsymbol{v}\!+\!\left(\!\frac{1}{2}\boldsymbol{v}^{2}\!-\! h\!+\! K\!\right)\! d\rho\nonumber \\
\!\!\!\!\!\!\!\!\!\! &  & \!+\!\left\{ \!\!\left(\!\frac{\partial\beta_{i}}{\partial t}\!+\!\partial_{j}(\beta_{i}v_{j})\!\!\right)\!-\!\left(\!\rho\partial_{j}K\!-\!\partial_{k}\sigma_{jk}\!-\! T\rho\partial_{j}s\!\right)\!\frac{\partial x_{j}}{\partial A_{i}}\!\right\} \! dA_{i}\nonumber \\
\!\!\!\!\!\!\!\!\!\! &  & +\left(\rho-\rho_{{\rm init}}J^{-1}\right)dK+(\partial_{t}A_{i}+\boldsymbol{v}\cdot\nabla A_{i})d\beta_{i}\nonumber \\
\!\!\!\!\!\!\!\!\!\! &  & \left.+\partial_{k}\!\!\left\{ \!\!\left(\!\!\beta_{i}v_{k}\!+\!\sigma_{jk}\frac{\partial x_{j}}{\partial A_{i}}\!+\!\rho K\frac{\partial x_{k}}{\partial A_{i}}\!\!\right)dA_{i}\!\!\right\} \!\right],\label{eq:act}
\end{eqnarray}
with the aid of Eq.~(\ref{eq:deltaS}). The last line in Eq.~(\ref{eq:act})
is the surface term yielding a boundary condition. Then, we obtain
Eqs.~(\ref{eq:Clebsch}), (\ref{eq:massCon}) and the followings,
\begin{eqnarray}
\!\!\!\!\!\!\!\!\!\!\!\! dv & \!\!:\!\! & \boldsymbol{v}+\frac{\beta_{j}}{\rho}\nabla A_{j}=\boldsymbol{0},\label{eq:velocity}\\
\!\!\!\!\!\!\!\!\!\!\!\! d\rho & \!\!:\!\! & \frac{1}{2}\boldsymbol{v}^{2}-h+K=0,\label{eq:K}\\
\!\!\!\!\!\!\!\!\!\!\!\! dA_{i} & \!\!:\!\! & \rho D_{t}\!\!\left(\!\frac{\text{\ensuremath{\beta_{i}}}}{\rho}\!\right)\!-\!\left(\rho\partial_{j}K\!-\!\partial_{k}\sigma_{jk}\,\!\!-\!\rho T\partial_{j}s\right)\!\!\frac{\partial x_{j}}{\partial A_{i}}\!\!=\!\!0.\label{eq:beta-1}
\end{eqnarray}
Here $h$ is enthalpy, defined as $\epsilon+p/\rho$. The Lie derivative
$L_{v}^{1}$ of Eq.~(\ref{eq:velocity}) yields
\begin{equation}
\partial_{t}\boldsymbol{v}+\nabla\boldsymbol{v}^{2}-\boldsymbol{v}\times(\nabla\times\boldsymbol{v})+D_{t}\left(\frac{\beta_{i}}{\rho}\right)\nabla A_{i}=\boldsymbol{0},\label{eq:Lv-1}
\end{equation}
with the aid of Eq.~(\ref{eq:Clebsch})\cite{Fukagawa01092010,bernard1984geometrical}.
Substituting Eq.~(\ref{eq:beta-1}) into Eq.~(\ref{eq:Lv-1}), we
obtain the equation of motion,
\begin{equation}
\frac{\partial}{\partial t}\left(\rho v_{j}\right)+\partial_{k}\left(\rho v_{j}v_{k}+P\delta_{jk}+\sigma_{jk}\right)=0.\label{eq:momentum-1}
\end{equation}
Using the same procedure in Sec.~\ref{sub:Noether's-theorem-with},
we obtain
\begin{eqnarray}
\!\!\!\!\!\!\!\!\!\! &  & \!\frac{\partial}{\partial t}\!\left(\!\beta_{i}\partial_{t}A_{i}\!-\!{\cal L}\!\right)\!+\!\partial_{j}\!\left\{ \!\beta_{i}v_{j}\frac{\partial A_{i}}{\partial t}\!+\!(\!\rho K\delta_{jk}\!-\!\sigma_{jk}\!)\!\frac{\partial x_{k}}{\partial A_{i}}\!\partial_{t}\! A_{i}\!\right\} \nonumber \\
\!\!\!\!\!\!\!\!\!\! &  & +Q=0,\label{eq:N}
\end{eqnarray}
and Eq.~(\ref{eq:EnergyCON}) with the aid of Eq.~(\ref{eq:act}).

\section{The equation for vaporization\label{sec:GAS-LIQUID-1}}

The action is given by the integral of the Lagrangian (\ref{eq:Le-1})
over the considered time and space. The stationary condition of the
action with subject to Eqs.~(\ref{eq:Clebsch}), (\ref{eq:massCon}),
and (\ref{eq:deltaS-1}) yields the sum of Eq.~(\ref{eq:act}) and
\begin{eqnarray}
\!\!\!\!\!\!\!\!\!\! &  & \left\{ \!\!-\frac{\partial E}{\partial\rho}+T\partial_{j}\left(\frac{1}{T}\frac{\partial E}{\partial\partial_{j}\rho}\right)\!\!\right\} d\rho\!\!-\!\!\frac{\left(\partial_{j}T\right)}{T}\frac{\partial E}{\partial\partial_{j}\rho}(\partial_{k}\rho)dX_{k}\nonumber \\
\!\!\!\!\!\!\!\!\!\! &  & +\partial_{j}\left\{ \frac{\partial E}{\partial\partial_{j}\rho}(\partial_{k}\rho)dX_{k}\right\} ,
\end{eqnarray}
Thus the optimized trajectory satisfies Eqs.~(\ref{eq:Clebsch}),
(\ref{eq:massCon}), (\ref{eq:velocity}), and the followings
\begin{eqnarray}
\!\!\!\!\!\!\!\!\!\! d\rho\!\!\!\! & : & \!\!\!\!\frac{1}{2}\boldsymbol{v}^{2}\!-\! h\!-\!\frac{\partial E}{\partial\rho}\!+\! T\nabla\!\cdot\!\left(\!\frac{1}{T}\frac{\partial E}{\partial\nabla\rho}\!\right)\!+\! K\!=\!0,\\
\!\!\!\!\!\!\!\!\!\! dA_{i}\!\!\!\! & : & \!\!\!\!\rho D_{t}\left(\frac{\text{\ensuremath{\beta_{i}}}}{\rho}\right)-\left\{ \rho(\partial_{j}K-T\partial_{j}s)-\partial_{k}\sigma{}_{jk}\right.\nonumber \\
 &  & \left.-\frac{\left(\partial_{k}T\right)}{T}\frac{\partial E}{\partial\partial_{j}\rho}\partial_{k}\rho\right\} \frac{\partial x_{j}}{\partial A_{i}}=0.
\end{eqnarray}
Then we have Eq.~(\ref{eq:momentum}) with the aid of the identity
\begin{equation}
\left\{ \frac{\partial E}{\partial\rho}-\partial_{j}\left(\frac{\partial E}{\partial\partial_{j}\rho}\right)\right\} \partial_{k}\rho=\partial_{j}\left(E\delta_{jk}-\frac{\partial E}{\partial\partial_{j}\rho}\partial_{k}\rho\right).
\end{equation}

\section{The equation for dissolution\label{sec:Two-component-Fluid-1}}

We have $d\epsilon=-Pd\rho^{-1}+\mu d\psi+Tds$ in the thermodynamics.
Pressure $P$ and temperature $T$ are defined as same as Eq.~(\ref{eq:PT}).
The coefficient $\mu\equiv(\partial\epsilon/\partial\psi)_{s,\psi}$
is an appropriately defined chemical potential of mixture, $\mu=\mu_{{\rm solute}}/m_{{\rm solute}}-\mu_{{\rm solvent}}/m_{{\rm solvent}}$,
where $\mu_{{\rm solute}}$ and $\mu_{{\rm solvent}}$ are the chemical
potentials of the two substances, and $m_{{\rm solute}}$ and $m_{{\rm solvent}}$
are the masses of the two kinds of the particles as in Sec.~58 of
Ref.\,\cite{landau1959fm}. The kinetic energy density is given by
$\sum_{c}\frac{1}{2}\rho_{c}\boldsymbol{v}_{c}^{2},$ where $\boldsymbol{v}_{c}$
is defined as $\boldsymbol{v}_{c}=(\rho_{c}\boldsymbol{v}+\boldsymbol{j}_{c})/\rho$.
We can rewrite it into

\begin{equation}
\frac{1}{2}\rho\boldsymbol{v}^{2}+\frac{1}{2\rho}\left(\frac{1}{\psi}+\frac{1}{1-\psi}\right)\boldsymbol{j}^{2}.\label{eq:KE}
\end{equation}
Thus the total kinetic energy density (\ref{eq:KE}) is given by the
sum of the kinetic energies associated to the mass average velocity
and the diffusion fluxes of the solute and the solvent. The Lagrangian
density ${\cal L}$ is given by
\begin{equation}
{\cal L}\equiv\rho\frac{1}{2}\boldsymbol{v}^{2}+\frac{1}{2\rho}\left(\frac{1}{\psi}+\frac{1}{1-\psi}\right)\boldsymbol{j}^{2}-\left(\rho\epsilon+E\right).\label{eq:Le-1-1}
\end{equation}
We use the same procedure in Sec.~\ref{sec:GAS-LIQUID}. The Lagrangian
to be minimized is the sum of Eq.~(\ref{eq:action-4}) and 
\begin{equation}
\gamma(\rho D_{t}\psi+\nabla\cdot\boldsymbol{j})+b_{i}\left\{ (D_{t}a_{i})J^{-1}-\boldsymbol{j}\nabla A_{i}\right\} =0.,\label{eq:action-3-2}
\end{equation}
where $\gamma$ and $\boldsymbol{b}$ are undetermined multipliers.
Solving the stationary condition with subject to the nonholonomic
condition (\ref{eq:deltaS-1-1}) yields Eqs.~(\ref{eq:Clebsch}),
(\ref{eq:massCon}), (\ref{eq:velocity}), and the followings
\begin{eqnarray}
dv & : & \boldsymbol{v}+\frac{\beta_{j}}{\rho}\nabla A_{j}+\gamma\nabla\psi+\frac{b_{j}}{\rho_{{\rm init}}}\nabla a_{j}=\boldsymbol{0},\label{eq:velocity-2}\\
d\rho & : & \frac{1}{2}\boldsymbol{v}^{2}-h-\frac{\partial E}{\partial\rho}+K\nonumber \\
 &  & +\gamma D_{t}\psi-\frac{1}{2\rho^{2}}\left(\frac{1}{\psi}+\frac{1}{1-\psi}\right)\boldsymbol{j}^{2}=0,\label{eq:K-2}\\
dA_{i} & : & \rho D_{t}\left(\frac{\beta_{i}}{\rho}\right)-\left\{ \rho(\partial_{j}K-T\partial_{j}s)-\nu_{k}\partial_{j}a_{k}\right.\nonumber \\
 &  & \left.-\frac{\partial_{k}T}{T}\frac{\partial E}{\partial\partial_{j}\psi}\partial_{k}\psi-\partial_{k}\sigma{}_{jk}\right\} \frac{\partial x_{j}}{\partial A_{i}}=0,\label{eq:beta}\\
dj & : & \frac{1}{\rho}\left(\!\frac{1}{\psi}\!+\!\frac{1}{1-\psi}\!\right)\boldsymbol{j}-\!\nabla\gamma-b_{i}\nabla A_{i}=0,\label{eq:DiffusionFlux}\\
d\psi & : & \rho\mu+\frac{\partial E}{\partial\psi}-T\nabla\cdot\left(\frac{1}{T}\frac{\partial E}{\partial\nabla\psi}\right)+\rho D_{t}\gamma\nonumber \\
 &  & +\frac{1}{2\rho}\left(\frac{1}{\psi^{2}}-\frac{1}{(1-\psi)^{2}}\right)\boldsymbol{j}^{2}=0,\label{eq:etaalpha}\\
d\boldsymbol{a} & : & D_{t}\left(\frac{b_{i}}{\rho_{{\rm init}}}\right)+\frac{\nu_{i}}{\rho}=0\label{eq:q-3}
\end{eqnarray}
Here we use $J^{-1}=\rho/\rho_{{\rm init}}$ and Eq.~(\ref{eq:massCon}).
We have Eq.~(\ref{eq:momentum-2-1}) from Eqs.~(\ref{eq:velocity-2})--(\ref{eq:beta}),
and Eq.~(\ref{eq:DiffusionFlux-1}) from Eqs.~(\ref{eq:DiffusionFlux})--(\ref{eq:q-3}).

\section{The equation of a chiral LC\label{sec:NLCcal}}

The total kinetic energy destiny is given by the sum of momentum energy
$\rho\boldsymbol{v}^{2}/2$, and angular momentum energy per mass,
$\rho(\boldsymbol{\omega}{}^{t}I\boldsymbol{\omega})/2$. Here $I_{ij}$
is the molecular moment of inertia, and promotional to $\delta_{ij}$.
The total energy is given by the sum of the bulk internal energy $\rho\epsilon(\rho,s,n_{i},\partial_{j}n_{i})$
and the surface energy $E(\nabla\rho,n_{i},\partial_{j}n_{i})$. The
Lagrangian density is given by
\begin{equation}
{\cal L}\equiv\rho\left\{ \frac{1}{2}\boldsymbol{v}^{2}+\frac{1}{2}\boldsymbol{\omega}^{t}I\boldsymbol{\omega}-\epsilon\right\} -E,\label{eq:Le-1-2}
\end{equation}
which satisfies the time and space translation symmetries, and the
rotational symmetry. The modified Lagrangian density is given by the
sum of Eq.~(\ref{eq:action-4}) and 
\begin{equation}
\lambda_{i}(\partial_{t}n_{i}+(\boldsymbol{v}\cdot\nabla)n_{i}-\omega_{i})\label{eq:action-5}
\end{equation}
where ${\cal L}$ is Eq.~(\ref{eq:Le-1-2}) and $\lambda_{i}$ is
an undetermined multiplier. The stationary condition of Eq.~(\ref{eq:action-5})
with subject to the constraint (\ref{eq:deltaS-3}) regenerates Eqs.~(\ref{eq:Clebsch}),
(\ref{eq:massCon}), (\ref{eq:omega}), and yield
\begin{eqnarray}
d\boldsymbol{v} & : & \boldsymbol{v}+\frac{\beta_{j}}{\rho}\nabla A_{j}+\frac{\lambda_{j}}{\rho}\nabla n_{j}=\boldsymbol{0},\label{eq:v}\\
d\boldsymbol{\omega} & : & I_{ij}\omega_{j}-\frac{\lambda_{i}}{\rho}=0,\label{eq:DeltaOmega}\\
d\rho & : & \frac{1}{2}\boldsymbol{v}^{2}+\frac{1}{2}\boldsymbol{\omega}^{t}I\boldsymbol{\omega}-h+K\nonumber \\
 &  & -\frac{\partial E}{\partial\rho}+T\nabla\cdot\left(\frac{1}{T}\frac{\partial E}{\partial\nabla\rho}\right)=0,\label{eq:kappa}\\
dA_{i} & : & \rho D_{t}\left(\frac{\text{\ensuremath{\beta_{i}}}}{\rho}\right)-\Bigl\{\rho\partial_{j}K-\rho T\partial_{j}s\nonumber \\
 &  & \ \left.-\partial_{i}\sigma_{ij}-g_{i}\partial_{j}(\nabla n_{i})\right\} \frac{\partial x_{j}}{\partial A_{i}}=0,\label{eq:beta-1-2}\\
dn_{i} & : & \partial_{j}(\rho\pi_{ij})-\rho D_{t}\left(\frac{\lambda_{i}}{\rho}\right)+g_{i}=0.\label{eq:DeltaN}
\end{eqnarray}
Here, we define $l_{i}$ and $\pi_{ij}$ as
\begin{equation}
l_{i}\equiv\left(\frac{\partial\epsilon}{\partial n_{i}}\right)_{\rho,s}{\rm and}\ \pi_{ij}\equiv\left(\frac{\partial\epsilon}{\partial(\partial_{i}n_{j})}\right)_{\rho,s}.
\end{equation}
The equation (\ref{eq:beta-1-2}) becomes
\begin{eqnarray}
 &  & \rho D_{t}\left(\frac{\text{\ensuremath{\beta_{i}}}}{\rho}\right)+\left\{ \frac{\rho}{2}\partial_{j}(\boldsymbol{v}^{2}+\boldsymbol{\omega}^{t}I\boldsymbol{\omega})-\partial_{j}P+\partial_{k}\sigma_{jk}\right.\nonumber \\
 &  & -\rho\pi_{lm}\partial_{j}\partial_{l}n_{m}+g_{k}\partial_{j}n_{k}\Bigr\}\frac{\partial x_{j}}{\partial A_{i}}=0,\label{eq:beta-4}
\end{eqnarray}
with the aid of Eq.~(\ref{eq:kappa}) and 
\begin{eqnarray}
\partial_{j}h & = & \partial_{j}\left(\epsilon+\frac{P}{\rho}\right)\nonumber \\
 & = & \frac{1}{\rho}\partial_{j}P+T\partial_{j}s+\pi_{lm}\partial_{j}\partial_{l}n_{m},
\end{eqnarray}
We use $d\epsilon=-Pd\rho^{-1}+Tds+l_{i}dn_{i}+\pi_{ij}d\partial_{i}n_{j}$,
where $P$ and $T$ are given as same as Eq.~(\ref{eq:PT}). The
Lie derivative $L_{\boldsymbol{v}}^{1}$ of Eq.~(\ref{eq:v}) yields
\begin{equation}
L_{\boldsymbol{v}}^{1}\boldsymbol{v}+\left(D_{t}\frac{\beta_{i}}{\rho}\right)\nabla A_{i}+L_{\boldsymbol{v}}^{1}\left(\frac{\lambda_{i}}{\rho}\nabla n_{i}\right)=0,\label{eq:Dtv-2}
\end{equation}
with the aid of Eqs.~(\ref{eq:Clebsch}) and (\ref{eq:omega}). Using
Eqs.~(\ref{eq:DeltaOmega}) and (\ref{eq:DeltaN}), we can rewrite
$L_{\boldsymbol{v}}^{1}\left(\frac{\lambda_{i}}{\rho}\nabla n_{i}\right)$
into
\begin{eqnarray}
 &  & D_{t}\left(\frac{\lambda_{i}}{\rho}\right)\nabla n_{i}+\frac{\lambda_{i}}{\rho}\nabla(D_{t}n_{i})\nonumber \\
 & = & \frac{1}{\rho}\left(\partial_{j}(\rho\pi_{ij})+g_{i}\right)\nabla n_{i}-\frac{1}{2}\nabla(\boldsymbol{\omega}{}^{t}I\boldsymbol{\omega})\label{eq:ThidDtv}
\end{eqnarray}
Then we obtain Eq.~(\ref{eq:Vmotion}) from Eqs.~(\ref{eq:Dtv-2})
and (\ref{eq:ThidDtv}). The material derivative $D_{t}$ of Eq.~(\ref{eq:DeltaOmega})
yields the equation of the angular velocity $\boldsymbol{\omega}$
of a director,
\begin{equation}
\rho I_{ij}(\partial_{t}+\boldsymbol{v}\cdot\nabla)\omega_{j}-\partial_{j}\left(\rho\pi_{ij}+\frac{\partial E}{\partial(\partial_{i}n_{j})}\right)+l_{i}-\rho g_{i}=0.\label{eq:Nmotion}
\end{equation}
If we neglect the effect of surface energy $E$, the set of Eqs.~(\ref{eq:Vmotion})
and (\ref{eq:Nmotion}) is exactly the Ericksen-Leslie equations.
From Eqs.~(\ref{eq:nu-1}) and (\ref{eq:Nmotion}), the term $\eta\nabla T$
induce the Lehmann effect in a chiral LC, namely the rotation of the
director with the helical axis parallel to the heat current\cite{Leslie19111968,de1993physics,doi:10.1080/02678290902775281,PhysRevLett.100.217802,0295-5075-97-3-36006,0295-5075-83-1-16005,doi:10.1080/02678298908045691}.

\bibliographystyle{unsrt}
\bibliography{ref}

\begin{thebibliography}{10}

\bibitem{adkins1983equilibrium}
Clement~John Adkins.
\newblock {\em Equilibrium thermodynamics}.
\newblock Cambridge University Press, 1983.

\bibitem{JDvanderWaals}
J.D. van~der Waals.
\newblock The thermodynamic theory of capillarity flow under the hypothesis of
  a continuous variation in density.
\newblock {\em Verhandel. Konink. Akad. Weten. Amsterdam}, 1:1--56, 1893.

\bibitem{cahn1958free}
John~W Cahn and John~E Hilliard.
\newblock Free energy of a nonuniform system. i. interfacial free energy.
\newblock {\em The Journal of chemical physics}, 28(2):258--267, 1958.

\bibitem{doi:10.1143/JPSJ.78.052001}
Masao Doi.
\newblock Gel dynamics.
\newblock {\em Journal of the Physical Society of Japan}, 78(5):052001, 2009.

\bibitem{Lisin199755}
V.B. Lisin and A.I. Potapov.
\newblock Variational principle in the mechanics of liquid crystals.
\newblock {\em International Journal of Non-Linear Mechanics}, 32(1):55 -- 62,
  1997.

\bibitem{Lisin1999327}
V.B. Lisin and A.I. Potapov.
\newblock A variational method of deriving the equations of the non-linear
  mechanics of liquid crystals.
\newblock {\em Journal of Applied Mathematics and Mechanics}, 63(2):327 -- 332,
  1999.

\bibitem{PhysRevE.86.031703}
Yuri Obukhov, Tom\'as Ramos, and Guillermo Rubilar.
\newblock Relativistic lagrangian model of a nematic liquid crystal interacting
  with an electromagnetic field.
\newblock {\em Phys. Rev. E}, 86:031703, Sep 2012.

\bibitem{Leslie19111968}
F.~M. Leslie.
\newblock Some thermal effects in cholesteric liquid crystals.
\newblock {\em Proceedings of the Royal Society of London. Series A.
  Mathematical and Physical Sciences}, 307(1490):359--372, 1968.

\bibitem{de1993physics}
Pierre-Gilles De~Gennes and Jacques Prost.
\newblock {\em The physics of liquid crystals}, volume~23.
\newblock Clarendon press Oxford, 1993.

\bibitem{Fukagawa01052012}
Hiroki Fukagawa and Youhei Fujitani.
\newblock A variational principle for dissipative fluid dynamics.
\newblock {\em Progress of Theoretical Physics}, 127(5):921--935, 2012.

\bibitem{Fukagawa01092010}
Hiroki Fukagawa and Youhei Fujitani.
\newblock Clebsch potentials in the variational principle for a perfect fluid.
\newblock {\em Progress of Theoretical Physics}, 124(3):517--531, 2010.

\bibitem{fukagawa2012}
Hiroki Fukagawa.
\newblock {\em Improvements in the Variational Principle for Fluid Dynamics (In
  Japanese)}.
\newblock PhD thesis, Keio University, 2012.
\newblock In Japanese.

\bibitem{hyon2010energetic}
Yunkyong Hyon, Do~Young Kwak, and Chun Liu.
\newblock Energetic variational approach in complex fluids: maximum dissipation
  principle.
\newblock {\em DCDS-A}, 24(4):1291--1304, 2010.

\bibitem{liu2009introduction}
Chun Liu.
\newblock {\em An introduction of elastic complex fluids: an energetic
  variational approach}.
\newblock World Scientific: Singapore, 2009.

\bibitem{2014arXiv1407.1035J}
B.~D. {Jones}.
\newblock {Navier-Stokes Hamiltonian}.
\newblock {\em ArXiv e-prints}, July 2014.

\bibitem{aoki2014constraint}
Kunihiro Aoki.
\newblock A constraint on the thickness-weighted average equation of motion
  deduced from energetics.
\newblock {\em Journal of Marine Research}, 72(5):355--382, 2014.

\bibitem{bloch2003nonholonomic}
Anthony~M Bloch.
\newblock {\em Nonholonomic mechanics and control}, volume~24.
\newblock Springer, 2003.

\bibitem{pontryagin1987mathematical}
Lev~Semenovich Pontryagin.
\newblock {\em Mathematical theory of optimal processes}.
\newblock CRC Press, 1987.

\bibitem{free}
Toshihiro Iwai.
\newblock A geometric setting for classical molecular dynamics.
\newblock 1987.

\bibitem{Suzuki20111904}
Masuo Suzuki.
\newblock Irreversibility and entropy production in transport phenomena i.
\newblock {\em Physica A: Statistical Mechanics and its Applications},
  390(11):1904 -- 1916, 2011.

\bibitem{Suzuki20121074}
Masuo Suzuki.
\newblock Irreversibility and entropy production in transport phenomena, ii:
  Statistical mechanical theory on steady states including thermal disturbance
  and energy supply.
\newblock {\em Physica A: Statistical Mechanics and its Applications},
  391(4):1074 -- 1086, 2012.

\bibitem{Suzuki2013314}
Masuo Suzuki.
\newblock Irreversibility and entropy production in transport phenomena, iii:
  Principle of minimum integrated entropy production including nonlinear
  responses.
\newblock {\em Physica A: Statistical Mechanics and its Applications},
  392(2):314 -- 325, 2013.

\bibitem{Suzuki20134279}
Masuo Suzuki.
\newblock Irreversibility and entropy production in transport phenomena, iv:
  Symmetry, integrated intermediate processes and separated variational
  principles for multi-currents.
\newblock {\em Physica A: Statistical Mechanics and its Applications},
  392(19):4279 -- 4287, 2013.

\bibitem{PhysRev.37.405}
Lars Onsager.
\newblock Reciprocal relations in irreversible processes. i.
\newblock {\em Phys. Rev.}, 37:405--426, Feb 1931.

\bibitem{Kambe200798}
Tsutomu Kambe.
\newblock Gauge principle and variational formulation for ideal fluids with
  reference to translation symmetry.
\newblock {\em Fluid Dynamics Research}, 39(1):98 -- 120, 2007.
\newblock In memoriam: Professor Isao Imai, 1914-2004.

\bibitem{israelachvili2011intermolecular}
Jacob~N Israelachvili.
\newblock {\em Intermolecular and surface forces: revised third edition}.
\newblock Academic press, 2011.

\bibitem{0295-5075-84-3-36003}
R.~Teshigawara and A.~Onuki.
\newblock Droplet evaporation in one-component fluids: Dynamic van der waals
  theory.
\newblock {\em EPL (Europhysics Letters)}, 84(3):36003, 2008.

\bibitem{PhysRevE.75.03630s4}
Akira Onuki.
\newblock Dynamic van der waals theory.
\newblock {\em Phys. Rev. E}, 75:036304, Mar 2007.

\bibitem{landau1959fm}
L.~D. Landau, E.~M. Lifshitz, J.~B. Sykes, and W.~H. Reid.
\newblock {\em Fluid Mechanics}.
\newblock Pergamon Press Oxford, England, 1959.

\bibitem{bernard1984geometrical}
Bernard~F. Schutz.
\newblock {\em Geometrical methods of mathematical physics}.
\newblock Cambridge University Press, 1984.

\bibitem{C4SM00670D}
Jun Yoshioka, Fumiya Ito, Yuto Suzuki, Hiroaki Takahashi, Hideaki Takizawa, and
  Yuka Tabe.
\newblock Director/barycentric rotation in cholesteric droplets under
  temperature gradient.
\newblock {\em Soft Matter}, 10:5869--5877, 2014.

\bibitem{:/content/aip/journal/pof1/9/6/10.1063/1.1761821}
J.~L. Ericksen.
\newblock Inequalities in liquid crystal theory.
\newblock {\em Physics of Fluids}, 9(6), 1966.

\bibitem{doi:10.1080/02678290902775281}
P.~Oswald and A.~Dequidt.
\newblock Lehmann effect in chiral liquid crystals and langmuir monolayers: an
  experimental survey.
\newblock {\em Liquid Crystals}, 36(10-11):1071--1084, 2009.

\bibitem{PhysRevLett.100.217802}
Patrick Oswald and Alain Dequidt.
\newblock Measurement of the continuous lehmann rotation of cholesteric
  droplets subjected to a temperature gradient.
\newblock {\em Phys. Rev. Lett.}, 100:217802, May 2008.

\bibitem{0295-5075-97-3-36006}
P.~Oswald.
\newblock About the leslie explanation of the lehmann effect in cholesteric
  liquid crystals.
\newblock {\em EPL (Europhysics Letters)}, 97(3):36006, 2012.

\bibitem{0295-5075-83-1-16005}
P.~Oswald and A.~Dequidt.
\newblock Direct measurement of the thermomechanical lehmann coefficient in a
  compensated cholesteric liquid crystal.
\newblock {\em EPL (Europhysics Letters)}, 83(1):16005, 2008.

\bibitem{doi:10.1080/02678298908045691}
N.~V. Madhusudana and R.~Pratibha.
\newblock An experimental investigation of electromechanical coupling in
  cholesteric liquid crystals.
\newblock {\em Liquid Crystals}, 5(6):1827--1840, 1989.

\end{thebibliography}

\end{document}